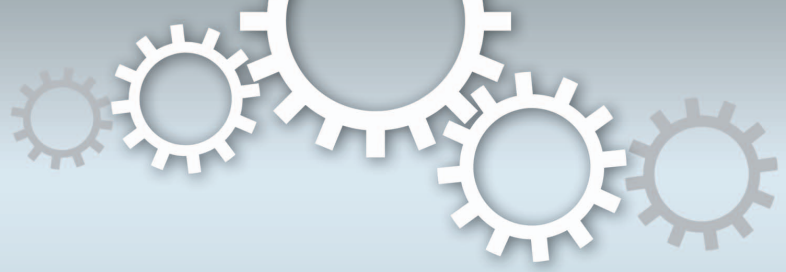

**OPEN**

# High-density Two-Dimensional Small Polaron Gas in a Delta-Doped Mott Insulator

Daniel G. Ouellette[1], Pouya Moetakef[2], Tyler A. Cain[2], Jack Y. Zhang[2], Susanne Stemmer[2], David Emin[3] & S. James Allen[1]

[1]Department of Physics, University of California, Santa Barbara, California, [2]Materials Department, University of California, Santa Barbara, California, [3]Department of Physics and Astronomy, University of New Mexico, Albuquerque, New Mexico.

Correspondence and requests for materials should be addressed to S.J.A. (allen@itst.ucsb.edu)

Heterointerfaces in complex oxide systems open new arenas in which to test models of strongly correlated material, explore the role of dimensionality in metal-insulator-transitions (MITs) and small polaron formation. Close to the quantum critical point Mott MITs depend on band filling controlled by random disordered substitutional doping. Delta-doped Mott insulators are potentially free of random disorder and introduce a new arena in which to explore the effect of electron correlations and dimensionality. Epitaxial films of the prototypical Mott insulator $GdTiO_3$ are delta-doped by substituting a single $(GdO)^{+1}$ plane with a monolayer of charge neutral SrO to produce a two-dimensional system with high planar doping density. Unlike metallic $SrTiO_3$ quantum wells in $GdTiO_3$ the single SrO delta-doped layer exhibits thermally activated DC and optical conductivity that agree in a quantitative manner with predictions of small polaron transport but with an extremely high two-dimensional density of polarons, $\sim 7 \times 10^{14}$ cm$^{-2}$.

H eterointerfaces in complex oxide systems open new arenas in which to test models of strongly correlated material, explore the role of dimensionality in metal-insulator-transitions (MITs) and small polaron formation, and consider technological applications made possible by interface charge density control[1]. Many transition metal oxides, particularly those close to the quantum critical point, exhibit Mott MITs that are dependent on band filling[2]. In the bulk, these MITs are typically controlled by disordered (random) substitutional doping[3–5].

The atomic-scale precision of oxide molecular beam epitaxy (MBE) allows growth of a single, nominally continuous, plane of dopants within an epitaxial film of a Mott insulating oxide. The delta-doping of a complex oxide should be distinguished from delta-doping of conventional semiconductors. The latter are well described by carriers that can be described within an effective mass approximation, bound to the dopant plane by a self-consistent electrostatic potential, but that spill over many lattice planes. Electron correlations do not change in a qualitative way one–electron models of these electron states, their transport, and optical properties. We note that delta-doping $LaTiO_3$ in $SrTiO_3$[6–8] is qualitatively similar to the semiconductor case, the difference being the band nature of the host $SrTiO_3$. In particular, delta-doping the band insulator $SrTiO_3$ with a narrow sheet of La donors or one complete formula unit of $LaTiO_3$ leads to a metallic 2D electron gas characterized by several occupied sub-bands[8]. Unlike the aforementioned semiconductor analogs, the electron states in the one layer thick delta-doping layer in a Mott band insulator defy effective mass treatments and will be controlled by local bonding rearrangements, lattice distortions[9], and strong electron-electron interactions. Of particular interest are Mott materials close to the MIT where the effective doping level may be sufficient to drive the host system into the metallic state. Equally important, unlike control of the MIT in the bulk by random alloy doping, disorder is minimized with layer-by-layer MBE growth of the host and the doping layer. We expect a two-dimensional electron/hole system to emerge with properties that may be distinct from both the host and bulk alloy doping.

Previous work by us on $SrTiO_3$ quantum wells in $GdTiO_3$ has shown that $\sim 7 \times 10^{14}$ cm$^{-2}$ electrons are confined between two $GdTiO_3$ barriers[10]. This invites exploration of the electron states, transport and optical absorption in ultra-thin quantum wells at the extreme delta-doped limit, a single $(SrO)^0$ layer substituted for a $(GdO)^{+1}$ layer in $GdTiO_3$ host. $GdTiO_3$ is one of the prototypical Mott insulating rare-earth ($R$) titanate perovskites, $RTiO_3$, which exhibit a bandwidth and filling controlled MIT[4,5]. An SrO layer in $GdTiO_3$ replaces $Gd^{3+}$ with $Sr^{2+}$ in a single atomic plane and introduces one hole per planar unit cell. This may be viewed as the extreme limit of the aforementioned SrTiO$_3$ quantum well with ½ an electron or ½ a hole per Ti layer on either side of the SrO plane in the neighboring GdTiO$_3$ insulator. From the latter perspective, if one assumes the holes are restricted to the TiO$_2$ layers neighboring the SrO layer, the electron concentration per Ti[5,11] is far from that needed for a Mott insulating state. In the bulk, doping at 50% brings GdTiO$_3$ well into the metallic phase. However, recent *ab initio* and model calculations based on this structure have predicted that a two-dimensional Mott-insulating state may persist if the electron states of the two TiO$_2$ planes on either side of the SrO layer dimerize[12]. Furthermore, structures with 2 SrO layers embedded in GdTiO$_3$ were found to be insulating, exhibiting activated transport, while structures with 3 SrO layers, albeit still metallic, exhibited signatures of mass enhancement in DC transport[13].

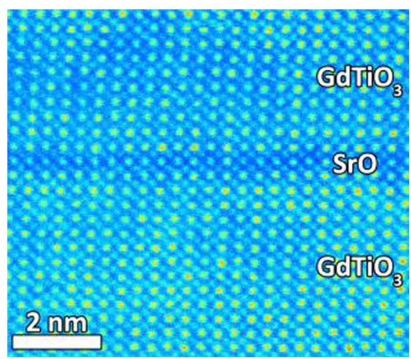

Figure 1 | TEM micrograph of a single SrO layer in GdTiO$_3$.

To probe the ground state and dynamics of this delta-doped Mott insulator, we measure the resistivity, Hall effect, thermopower and optical conductivity. These properties display a strong correspondence to the theory of transport by non-interacting small polarons[14,15]. We focus in particular on the broad resonant optical conductivity and its relation to the thermally activated DC transport while drawing on microscopic parameters for SrTiO$_3$ that are eminently reasonable to show that the system may be quantitatively described as a high-density, two-dimensional gas of small polarons.

## Results

Figure 1 shows a high-angle annular dark-field scanning transmission electron microscope micrograph showing a single SrO layer in GdTiO$_3$. Delta-doping with a plane of SrO results in an insulating state, despite the high concentration of holes. The effective DC conductivity $\sigma = 1/R_s d$, where $R_s$ is the sheet resistance and $d$ is the total film thickness, is shown in Fig. 2(a). At high temperatures the conductivity can be described by an Arrhenius function $\sigma(T) = \sigma_0 \exp\left(-E_a^A/kT\right)$. (The superscript $A$ distinguishes the activation energy for a simple Arrhenius function with constant prefactor.) The activation energies (Table 1) for the delta-doped samples are moderately smaller than the undoped films but the magnitude of the conductivity (the prefactor) is substantially larger.

The high resistivity of the samples limited the measurements of the Hall effect and Seebeck coefficient to temperatures around room temperature. In Fig. 2(b) only sample B2 exhibited a measureable Hall effect. The Hall constant measured over a relatively narrow temperature range, is *electron-like*, but may be characterized by an activation energy comparable to that measured by the conductivity. The Seebeck coefficients measured at room temperature are *hole-like* and exhibited values of +22(5), +73(2), +153(2), and +24(5) for samples B2, B1, A3 and A1 respectively. They are comparable to values for bulk doped GdTiO$_3$ reported elsewhere[11].

The volume averaged optical conductivity of the GdTiO$_3$ and delta doped sample B2 are shown in Fig. 3(a). The optical conductivity of the pure GdTiO$_3$ films, which is widely attributed to the Mott-Hubbard gap[16], is similar to that of bulk titanates[3,16,17] and shows an onset near 0.5 eV. Delta-doping results in additional mid-gap absorption from incoherent processes with $\sigma(\omega)$ approaching at low frequencies the relatively small DC value. There is no coherent Drude response. With increased temperature, the absorption shifts to lower frequencies. In all samples, an optical phonon mode associated with Ti-O stretching is seen as a sharp peak at 67 meV.

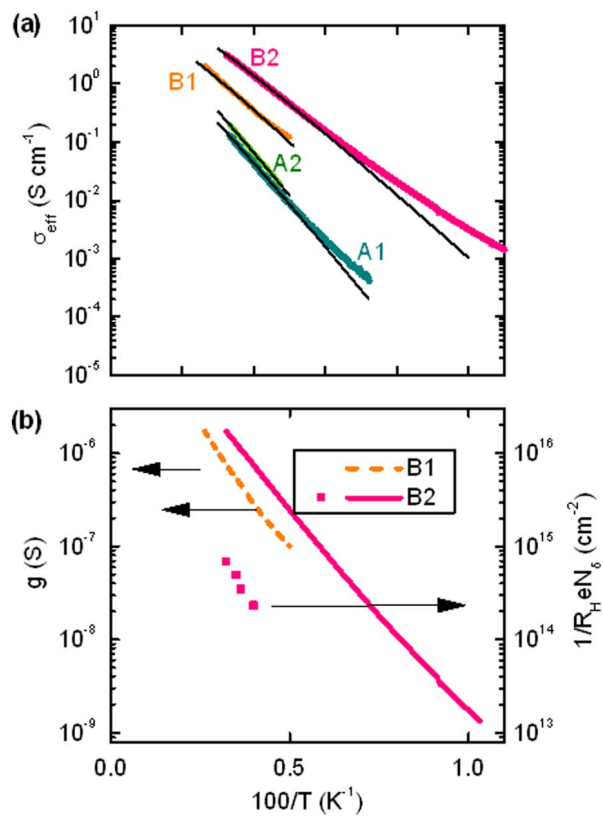

Figure 2 | (a) Fat colored lines: effective 3-dimesional conductivity versus inverse temperature. Thin black lines: fits to small polaron conductivity. (b) Lines: sheet conductivity per delta-doping plane (left axis) for delta-doped samples B1 and B2. Data points: Hall density per delta-doping plane on the same logarithmic scale (right axis).

In the region of the Mott-Hubbard gap, we can attempt to separate the frequency dependent conductivity of the delta-doped layer from the parallel contribution of the remaining GdTiO$_3$. The optical conductivity of GdTiO$_3$, $\sigma_{GTO}(\omega)$, is taken as the average of the measured response of samples A1 and A2 and is subtracted from the measured optical conductivity of sample B2, $\sigma_{B2}(\omega)$, to get the 2D conductivity $g(\omega)$ of a delta-doped layer,

$$g(\omega) = \frac{d}{N_\delta}\left(\sigma_{B2}(\omega) - \frac{d_{GdTiO_3}}{d}\sigma_{GdTiO_3}(\omega)\right) \quad (1)$$

where $N_\delta = 4$ is the number of delta-doped layers in the superlattice and $d_{GTO}$ is the thickness of GdTiO$_3$ contributing a bulk-like optical conductivity. This calculation hinges on an estimate of the unknown spatial extent of the delta-doped carriers. To avoid ambiguity we take $d_{GTO}/d = 1$ and obtain the $g(\omega)$, shown in Fig. 3(b) as a lower bound. It exhibits a broad peak centered around 0.5–0.8 eV. The dc sheet conductance of the delta doped layer(s) shown in Fig. 2(b) is corrected in a similar manner by subtraction of the (comparatively small) dc conductivity of the equivalent thickness of GdTiO$_3$.

Table 1 | Sample parameters and transport properties. From left to right: sample name, total film thickness d, number of delta-doped layers $N_\delta$, the number of SrO layers $t_{STO}$, room temperature resistivity ρ (300 K), Arrhenius activation energy $E_a^A$ for conductivity, small polaron transport activation energy $E_a$, measured prefactor, and calculated small polaron prefactor using parameters consistent with the optical conductivity. Samples A1–A3 are nominally undoped $GdTiO_3$ films. The activation energies are from fits over the range ~200–400 K. Data for sample A3 and A1 have been published previously[33,11]. No transport data is available for sample A2 although it was used to obtain optical conductivity for the nominally undoped films. B1 and B2 are $GdTiO_3$ films delta doped with single layers of SrO. B1 has a single layer while B2 has 4 such layers separated from each other by ~4 nm of $GdTiO_3$

| (1) Sample | (2) d(nm) | (3) $N_\delta$ | (4) $t_{STO}$ | (5) ρ (300 K) (Ω cm) | (6) $E_a^A$ (meV) $\rho_0 \exp\left(\frac{E_a^A}{k_B T}\right)$ | (7) $E_a$(fit) (meV) $\sigma = \frac{C}{k_B T} \exp\left(-\frac{E_a}{k_B T}\right)$ | (8) C (fit) (S/cm)J | (9) $C = ne^2 a^2 \omega_0/(2\pi)$ (S/cm)J |
|---|---|---|---|---|---|---|---|---|
| A1 | 10 | 0 | | 7.2 | 129 ± 5 | 160 | 2.6 $10^{-19}$ | |
| A2 | 19.5 | 0 | | | | | | |
| A3 | 11 | 0 | | 5.2 | 144 ± 5 | 162 | 4.8 $10^{-19}$ | |
| B1 | 8 | 1 | 1 | 1.24 | 106 ± 10 | 130 ± 10 | 5.0 $10^{-19}$ | 8.45 $10^{-19}$ |
| B2 | 21.5 | 4 | 1 | 0.35 | 93 ± 5 | 111 ± 5 | 1.0 $10^{-18}$ | 1.25 $10^{-18}$ |

The spectral weight is quantified according to,

$$n_{3D}(\omega) = \frac{2m_e}{e^2\pi} \int_0^\omega \sigma_1(\omega') d\omega' \quad (2)$$

where $m_e$ is the free electron mass. The results for $GdTiO_3$ films A1 and A2 and the delta-doped superlattice B2 are shown in Fig. 3(c). The difference is attributed to the delta-doping and is plotted using the right axis in units of a two-dimensional density per delta doped plane according to $n_{2D} = d n_{3D}/N_\delta$. The resulting $n_{2D}$ is substantially less (~1/10) than the nominal doping density ~$7 \times 10^{14}$ associated with each SrO plane. By comparison, the modest suppression of the spectral weight of *metallic* two[6] and three[18]-dimensional electron gases in $SrTiO_3$ is attributed to an approximately three-fold enhancement over the bare band mass in equation (2) to ~$2m_e$[19].

## Discussion

$GdTiO_3$ is a Mott insulator. Beyond a critical doping, $x \sim 0.2$[11,20], the system $Gd_{1-x}Sr_xTiO_3$ undergoes a MIT to a metal with carrier density significantly larger than the doping concentration. We find that delta doping far exceeding the critical concentration for $GdTiO_3$ does not render it metallic. Rather, as evidenced by optical and electronic transport measurements discussed in the following, the system remains an insulator; small polaron formation occurs when substituting an SrO plane for a GdO plane confining the resulting holes to the adjacent ($TiO_2$) planes[21–23].

Our optical measurements do not detect the characteristic Drude response from mobile charge carriers. Rather, we observe the broad carrier-induced absorption band characteristic of small polarons. This absorption primarily transfers a small polaron's electronic carrier from the site at which it is self-trapped to a neighboring site. Near its peak the absorption band's profile is a modified Gaussian centered at the difference between the final site's electronic energy and that of the self-trapped carrier, $2E_b$[15,24,25]

$$\sigma_1(\omega) \approx n_p \frac{2\pi^{1/2} e^2 a^2 t^2}{\omega \hbar^2 \Delta} \exp\left[-\frac{(2E_b - \hbar\omega)^2}{\Delta^2}\right], \quad (3)$$

where $E_b$ is the polaron binding energy.

The absorption strength is determined by the polaron density, $n_p$, the electronic transfer energy $t$ and the inter-site separation, $a$. At low temperatures $kT \ll \hbar\omega_o$ the width is determined by atoms' zero-point vibrations, $\Delta = (4E_b\hbar\omega_o)^{1/2}$, where $\hbar\omega_o$ is the energy of a representative phonon mode. The sum rule for this absorption process, Eq. (2), yields $\sim (n_p e^2/m_e)(t/E_b)$.

Equation (3) is fit to the optical conductivity of the delta-doped layer in Fig. 3(b). We assume a hole density ~$7 \times 10^{14}$ cm$^2$, the nominal 2D doping. The fit at 10 K (296 K) gives $E_b$ = 440 (370) meV, $t$ = 110 (100) meV, $\Delta$ = 440 (402) meV, assuming $\hbar\omega_0$ = 110 meV. The deduced transfer energy $t$ is comparable to that obtained from first-principles calculations[12]. The summing of the small-polaron conductivity described by Eq. (2) is shown in Fig. 3(c). Its magnitude implies that $t$ is significantly smaller than $E_b$, the condition for small-polaron formation.

Recent work by Zhang et al.[9] is particularly relevant here. They show that the thinnest $SrTiO_3$ quantum wells suffer interface distortions that will reduce $t$. Small transfer energy $t$ favors localization by either electron correlation or small polaron formation.

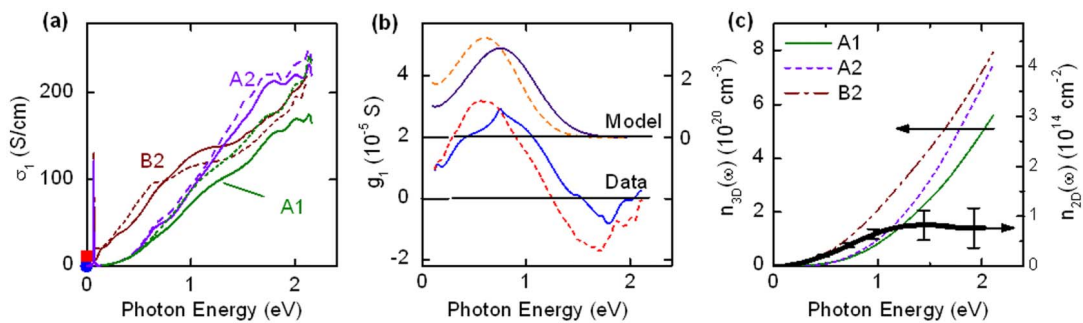

**Figure 3** | (a) Volume averaged 3D optical conductivity at 10 K (solid) and 296 K (dashed) for the indicated samples. Symbols indicate the dc conductivity 10 K (blue circle) and 296 K (red square). (b) Two-dimensional optical conductivity of the delta-doped layer after normalization together with the small polaron model, displaced upward for clarity. (see text) (c) Thin lines: sum rule obtained by integrating the (3D) optical conductivity according to (1). Fat line: delta doping contribution to sum rule. The right axis expresses the difference in units of 2D density per delta-doping plane. The error bars were estimated from the uncertainty in the reflectance measurement.

The dc conductivities of our films are shown in Fig. 2(a). These thermally activated conductivities are consistent with high-temperature adiabatic hopping of a temperature-independent density of small polarons $\sigma = \frac{C}{k_B T} \exp\left(-\frac{E_a}{k_B T}\right)$ where $C = (N_\delta/(a^2 d))(a^2 e^2/h)(\hbar\omega_o)$. The fit value of $C$ compares well with the values predicted by the model. See the last two columns in Table 1. (The assumed concentration of polarons, $n_P = N_\delta/(a^2 d)$, is an important parameter here. By the same token, we have no estimate of the prefactors for the nominally undoped samples since we have no measure of the background doping.) The activation energy for adiabatic hopping in the Holstein model is $E_a = E_B/2 - t$, where $t$ is electronic transfer energy characterizing a hop[26]. The measured activation energies for the delta-doped films (~0.1 eV) are comparable to those predicted with the parameters obtained from the fits to the small-polaron optical absorption.

We attribute the activation energies (0.1–0.14 eV) of our films' conductivities to small polaron hopping; they are substantially smaller than the Mott-Hubbard gap of undoped GdTiO$_3$ determined by the onset of its optical absorption. The intentionally delta-doped films have significantly higher values of $\sigma_o$(300 K) than do the nominally undoped films. Presumably the total carrier concentration for the delta-doped films is larger, although the carriers are restricted to the planes near to the SrO doping layer. The activation energies of the delta-doped films (~0.1 eV) are smaller than the nominally undoped films (~0.14 eV). This is consistent with the notion that the in-plane two-dimensional polaron transport along delta-doped layers is not perturbed by the random doping that is present in the bulk alloy. By contrast, the activation energy for hopping of small-polarons introduced by random background dopants, is expected to be augmented by the energy required to free the carrier from the dopant.

The limited data on the Seebeck coefficient and Hall effect does not refine our quantitative model of a high density 2D polaron gas described above but is consistent with it. The observed $p$-type Seebeck coefficients are consistent with dopant-induced polarons hopping among Ti cations on the two TiO$_2$ layers that encompass the SrO doping layer as well as hopping related to the unintentionally doping in samples A1 and A3[11]. We also observe that the Hall constant $R_H$ of the B2 film is $n$-type. The activated behavior of $1/R_H$ shown in Fig. 2(b) is comparable to that of the conductivity. The closeness of the activation energies of the conductivity and $1/R_H$ implies that the Hall mobility, $\mu_H \equiv \sigma R_H$, depends only weakly on temperature. In the limited temperature range, the Hall mobility only varies from 0.015 to 0.019 cm$^2$/volt.sec. The low and weak temperature dependent Hall mobility is a general feature of small-polaron hopping[23,26–28]. Furthermore, the anomalous sign of the Hall coefficient, opposite to the sign of the Seebeck coefficient, is not unusual for hopping conduction and is well documented in other systems[29,30]. In particular, a combination of nearest-neighbor and next-nearest-neighbor hopping among cations on the two TiO$_2$ planar square lattices adjacent to the SrO doping layer can produce a Hall effect sign anomaly[28]. We do not rely on the thermopower and Hall effect measurements to quantitatively support the model but are satisfied at this point that they are consistent with it.

Optical absorption and thermally activated transport in a delta doped Mott insulator can be described in a quantitative manner by adiabatic small polaron hopping in an ordered two dimensional lattice. The small polaron model invokes the full two-dimensional density given by the doping introduced by the SrO layer. It is surprising that a single polaron model should quantitatively describe a two-dimensional polaron gas at such a high density, ~7 × 10$^{14}$ cm$^{-2}$. These transport experiments show that the electron or holes in the ground state of this system are "self-trapped" as small polarons.

Future experiments call for more extensive Hall and thermopower measurements of this unique 2-dimensional small polaron system. The energetics of the binding of the carriers to the dopant plane could be measured by electric subband optical and/or tunneling spectroscopy to constrain theories of the insulating state of this highly correlated 2D polaron system.

## Methods

Thin films of GdTiO$_3$ were grown using hybrid molecular beam epitaxy with elemental and metal-organic sources, detailed elsewhere[10,11,31]. The films were grown on (001) (LaAlO$_3$)$_{0.3}$(Sr$_2$AlTaO$_6$)$_{0.7}$ (LSAT) substrates using two different growth modes. The first was a shutter controlled method[31] that deposits alternating layers of GdO and TiO$_2$. The second was a co-deposition method[32]. The resulting films and GdTiO$_3$/SrTiO$_3$ multilayers had very similar transport and magnetic properties. Single layers of SrO were grown using rate calibrations for SrTiO$_3$. Figure 1 shows a high-angle annular dark-field scanning transmission electron microscope micrograph showing a single SrO layer in GdTiO$_3$. Continued exposure at these rates also produced coherently grown SrTiO$_3$ quantum wells, referred to above. High angle annular dark field scanning TEM images of similar samples consistently indicate abrupt interfaces and unit cell thickness control[9]. We recognize that it is highly unlikely that we have a continuous single SrO layer over the entire sample but assume that the sample fraction described by such is large enough to determine the transport and optical properties.

We focus primarily on the delta-doped samples but reference their transport to several other related films. The samples considered in this study are listed in Table 1 and include both undoped GdTiO$_3$ films and the delta-doped samples. Delta-doped sample B1 consists of a single SrO delta-doped layer surrounded on either side by 4 nm of GdTiO$_3$. A second sample, B2, is a 4 repeat superlattice with single SrO layers separated by 4 nm of GdTiO$_3$[11,33]. Only B2 with 4 repeats provides enough signal to extract the optical conductivity of the delta-doped layers.

Ohmic contacts were made by e-beam deposition of 50 nm Ti/300~nm Au. Magnetotransport measurements were performed in a Van der Pauw geometry using a Keithley source meter with samples mounted in Quantum Design Physical Properties Measurement Systems having 7 T or 14 T superconducting magnets, respectively. The high sample resistance limited the lowest temperatures at which resistivity could be measured. While the transverse magneto-resistance required for a measure of the Hall effect is inadvertently contaminated by the much stronger longitudinal magneto resistance, fitting to a second order polynomial from −14 to +14 Tesla allows one to extract the magnetoresistance which is an odd function of magnetic field and we use it as a measure of the Hall coefficient[34,35]. The sign was referenced to a known $n$-type sample. Using a standard technique, the Seebeck coefficient $S$ was extracted from linear fits of the thermal voltage versus the temperature difference $\Delta T$.

Infrared reflectance measurements (~10° incidence) were performed in a Bruker 66 v/S Fourier Transform IR (FTIR) spectrometer operated in vacuum. Samples were mounted on the cold finger of a helium flow cryostat enabling measurements over the temperature range 10–300 K. A translation stage and bellows built into the cryostat allow sample changes without breaking the vacuum environment of the spectrometer. Gold mirrors or bare LSAT substrates were used as references[36,37]. The optical conductivity was extracted from the reflectance by fitting to a multilayer solution to the electromagnetic boundary conditions with the complex dielectric function of each film modeled using a series of Lorentz oscillators[38]. The measured dielectric response of the LSAT substrate is included in the model. Since the infrared penetration depth is significantly greater than the film thickness, the measurement provides the effective volume averaged 3D dielectric response of the thin films. It is expected that the films grew oriented with the (110)$_O$ planes of the orthorhombic GdTiO$_3$ parallel to the LSAT (001)$_C$ surface[31]; as a result the measured optical conductivity is an appropriately weighted average of all three principle axes.

### Acknowledgments

D.G.O., S.S. and S.J.A. acknowledge funding through a DARPA program (W911NF-12-1-0574). P.M. was supported by NSF (DMR 1006640), T.A.C. by the Center for Energy Efficient Materials, an Energy Frontier Research Center funded by the DOE (Award No. DESC0001009) and by the Department of Defense through a NDSEG fellowship. J.Y.Z. by DOE (grant no. DEFG02-02ER45994) and a Department of Defense NDSEG fellowship. This work made use of NSF funded facilities in the UCSB MRL (DMR-1121053) as well as the UCSB Nanofabrication Facility, a part of the NSF-funded NNIN network. The authors thank David Awschalom for the use of transport facilities. The authors are very grateful for invaluable discussions with Ru Chen, Chuck-Hou Yee, Leon Balents, Lars Bjaalie, Anderson Janotti, Burak Himmetoglu, Chris Van-de-Walle, and Andrew Millis.

### Author contributions

D.O. performed optical measurements, analyzed the data and wrote manuscript text. D.O., P.M. and T.C. performed electrical measurements. P.M. and S.S. grew the epilayers. J.Z. and S.S. obtained T.E.M. images. D.E. developed theoretical models. S.A. performed electrical measurements and wrote manuscript text. All authors reviewed the manuscript.

### Additional information

**Competing financial interests:** The authors declare no competing financial interests.

**How to cite this article:** Ouellette, D.G. *et al.* High-density Two-Dimensional Small Polaron Gas in a Delta-Doped Mott Insulator. *Sci. Rep.* **3**, 3284; DOI:10.1038/srep03284 (2013).